\documentclass{llncs}

\usepackage{array}
\usepackage{graphicx}
\usepackage{amsmath}
\usepackage{amsfonts}
\usepackage{hyperref}
\usepackage{xcolor}
\usepackage{subcaption}
\usepackage{multirow}
\usepackage{microtype}
\usepackage{soul}
\usepackage{xcolor}

\graphicspath{{figures/}}

\title{Tractogram filtering\\ of anatomically non-plausible fibers\\ with geometric deep learning}

 \author{Pietro Astolfi\inst{1,2,3}\thanks{We gratefully acknowledge the support of NVIDIA Corporation with the donation of the Titan Xp GPU used for this research.}, Ruben Verhagen\inst{2}, Laurent Petit\inst{4}, Emanuele Olivetti\inst{1,2}, Jonathan Masci\inst{6}, Davide Boscaini\inst{5} and Paolo Avesani\inst{1,2}}
 
 \institute{NeuroInformatics Laboratory (NILab),
   Bruno Kessler Foundation, Trento, Italy
   \and
   Center for Mind and Brain Sciences (CIMeC),
   University of Trento, Italy
  \and
  PAVIS, Italian Institute of Technology, Genova, Italy
  \and
  GIN, IMN, CNRS, CEA, Université de Bordeaux, Bordeaux, France
  \and
  Technologies of Vision (TeV), Bruno Kessler Foundation, Trento, Italy
  \and
  NNAISENSE, Lugano, Switzerland\\
 \email{pastolfi@fbk.eu}\quad\email{avesani@fbk.eu}
 }

\begin{document}

\maketitle

\begin{abstract}

Tractograms are virtual representations of the white matter fibers of the brain. They are of primary interest for tasks like presurgical planning, and investigation of neuroplasticity or brain disorders. Each tractogram is composed of millions of fibers encoded as 3D polylines. Unfortunately, a large portion of those fibers are not anatomically plausible and can be considered artifacts of the tracking algorithms. Common methods for tractogram filtering are based on signal reconstruction, a principled approach, but unable to consider the knowledge of brain anatomy. In this work, we address the problem of tractogram filtering as a supervised learning problem by exploiting the ground truth annotations obtained with a recent heuristic method, which labels fibers as either anatomically plausible or non-plausible according to well-established anatomical properties. The intuitive idea is to model a fiber as a point cloud and the goal is to investigate whether and how a geometric deep learning model might capture its anatomical properties. Our contribution is an extension of the Dynamic Edge Convolution model that exploits the sequential relations of points in a fiber and discriminates with high accuracy plausible/non-plausible fibers.
\end{abstract}

\section{Introduction}%
\label{sec:introduction}

Tractography represents a powerful method to reconstruct the white matter fibers from diffusion magnetic resonance (MR) recordings \cite{basser_vivo_2000}. While this method provides a good approximation of the brain connectivity structure, there is an open issue of artifactual fibers \cite{maier-hein_challenge_2017,maier-hein_tractography-based_2016}, i.e. anatomically non-plausible pathways. We address the task of filtering out such artifactual fibers using a deep learning model.

The structural connectivity of the brain can be reconstructed from diffusion MR signal by a step of local estimation of the diffusivity model \cite{tournier_robust_2007} and a subsequent step of fiber tracking \cite{girard_towards_2014}. The outcome is a tractogram, a virtual representation of the axonal pathways in the white matter, where each fiber is encoded as 3D polylines, commonly referred to as \emph{streamlines}. Typically, a whole brain tractogram is composed of millions of streamlines.

Tractograms are providing valuable contributions to critical tasks like presurgical intervention planning, the detection of biomarkers for brain disorders and the investigation of neuroplasticity. For these purposes, the accuracy of fiber tracking is of paramount importance. While in the preprocessing of diffusion data it is common practice to denoise the signal before estimating the diffusivity model \cite{glasser_minimal_2013,zhuang_correction_2006}, there is no similar step after fiber tracking to filter out noisy streamlines.

The assessment of the accuracy of fiber tracking has been approached a few years ago with an open contest\footnote{\url{tractometer.org/ismrm\_2015\_challenge}}{} involving many (20) research groups. The contest was designed as a task of bundle detection on a dataset composed of simulated diffusion MRI data. The joint effort allowed the evaluation of the quality and limits of the most common tracking methods \cite{maier-hein_challenge_2017}.
The positive outcome has been the lack of false negative streamlines, the critical issue has concerned the many false positive errors, e.g. artifactual streamlines.

The occurrence of false positive streamlines is not surprising. The general strategy of tracking methods is to oversample the possible pathways to preserve the property of coverage of all true positive fibers \cite{rheault_bundle-specific_2019}. The tacit assumption is to postpone the task of filtering false positive streamlines to a subsequent post-processing step. The reason for this strategy is the difficulty of encoding anatomical priors into tracking algorithms.

The most common approach to reduce artifactual streamlines are methods based on the inverse problem of signal reconstruction, e.g. Life \cite{pestilli_evaluation_2014}, SIFT \cite{smith_sift:_2013}, Commit \cite{daducci_commit:_2015}. The intuitive idea is to estimate how much the orientation of a streamline explains the diffusion signal. Despite the principled criterion, these methods do not take into account the knowledge of brain anatomy, like another method based on the topological properties of streamlines~\cite{yeh_automatic_2019}.

A recent rule-based method, namely ExTractor \cite{laurent_petit_half_2019}, has been proposed to filter out artifactual streamlines from tractograms by following  anatomical principles. The rules that it proposes encode the geometrical and spatial properties of the streamlines with respect to the basics of white matter neuroanatomy in terms of association, projection, and commissural fibers. As output, ExTractor labels streamlines as \emph{anatomically plausible} or \emph{anatomically non-plausible}. 


In this work, we address for the first time the problem of tractogram filtering as a supervised learning problem. We need to train a binary classifier to discriminate between two classes: anatomically plausible (P) and non-plausible (nP) fibers. The intuitive idea is to exploit the labeling of fibers provided by the rule-based method \cite{laurent_petit_half_2019} and to adopt deep learning models to learn the features of streamlines underlying the rules. The ultimate goal is to have a fast run-time solution to filter large tractograms and a flexible method to transpose new expert annotations.


Given the sequential structure of a streamline, we have chosen as reference learning model a bidirectional LSTM (bLSTM) neural network \cite{huang_bidirectional_2015}. Although this model can exploit the sequential information, it requires forcing the streamline representation to a fixed number of points. For this reason, we consider Geometric Deep Learning (GDL) models \cite{masci_geometric_2016} that support more flexible and appropriate data representations. We investigate
a PointNet (PN) model \cite{qi_pointnet_2017}, where a streamline can be represented as a point cloud i.e. set of 3D points, and a Dynamic Edge Convolution (DEC) model \cite{wang_dynamic_2019}, which in addition to PN considers the relations between points belonging to the same spherical local context. Our experiments shows that both GDL models provide an improvement with respect to bLSTM.

Despite the better results provided by PN and DEC, these models are invariant to permutations of points in the input point cloud. It means that if we permute the points in a streamline classified as plausible, the model will continue to classify it as plausible albeit the altered sequence of points.
To overcome this issue we propose a Sequence Dynamic Edge Convolution (sDEC) model, an extension of the DEC model that introduces the property of being sequence sensitive.

While the sDEC model provides only a modest increase in accuracy with respect to PN in classifying P and nP fibers, the analysis of error distribution shows different behaviours. 
sDEC is more robust when fibers are long and curved: those are the type of fibers where PN performs worst and produces significantly more false positive errors.
In addition, visual inspection of the false negatives errors made by sDEC shows that these fibers look truly anatomically non-plausible.
We may argue that this apparent mismatch might be related to the noise of the labelling process.


\section{Materials}%
\label{sec:materials}

\paragraph{Dataset.} Our reference dataset is composed of a collection of tractograms of 20 subjects randomly selected from the HCP dataset \cite{van_essen_wu-minn_2013}, which is a publicly available curated MRI dataset. The tractograms were obtained computing (i) the constrained spherical deconvolution (CSD) model \cite{tournier_robust_2007} on the 3T DWI (1.25mm, 270 multishell gradients), and the (ii) Particle Filtering Anatomically Constrained Tractography (PF-ACT) algorithm \cite{girard_towards_2014}. Specifically, the tracking generated around $\sim 1$M fibers for each tractogram by seeding 16 points for each voxel with step size 0.5mm.  The tractogram of each subject has been non linearly coregistered to the MNI standard space, and, for computational purposes, all the fibers have been compressed to the most significant points~\cite{presseau_new_2015}.

\paragraph{ExTractor labelling.}
According to the premise of a supervised learning approach, we created a dataset where for each tractogram we labeled the fiber as either plausible (P) or non-plausible (nP). The procedure of labelling followed the heuristic rules defined by ExTractor~\cite{laurent_petit_half_2019}, a tool that encodes the current knowledge on the anatomical pathways of white matter structures. In particular, ExTractor carries out a 2-step procedure. First, fibers are marked as nP when they are either (i) shorter than $20$ mm, (ii) contain a loop, or (iii) are truncated, i.e., they terminate in the deep white matter. The second step is concerned with the labeling of fibers marked as P, by splitting the main pathways into the three macro classes: association, projection, and commissural. This selection of fibers is further refined to filter out non-plausible pathways using a clustering method \cite{cote_cleaning_2015}. The outcome of the ExTractor procedure is a balanced partition between P and nP fibers, irrespective of the different tracking algorithms and data sources~\cite{laurent_petit_half_2019}. On our dataset, Extractor resulted in $49.7\pm1.5\%$ of P streamlines on average on the 20 subjects, but more significantly it showed the presence of a massive percentage, $31.8\pm1.2\%$, of nP streamlines shorter than 20mm.


\section{Methods}%
\label{sec:methods}
The intuition underlying our work is to treat a streamline as an undirected sequence of 3D points, aiming to learn geometric and spatial features relevant for the tractography filtering task. Currently, the best way to achieve such a goal is by employing GDL models \cite{masci_geometric_2016}, which are designed to learn geometric features of graphs and point clouds. Among the existing GDL approaches, PointNet \cite{qi_pointnet_2017} is the most adopted method both for its simplicity and effectiveness. Nevertheless, such an approach does not consider relations between points. It learns global properties of a point cloud just by encoding all the points separately and then aggregating them in a single descriptor through global pooling. Conversely, Dynamic Edge Convolution (DEC) \cite{wang_dynamic_2019} considers the points relations, by encoding them as edges of a graph dynamically induced by the point cloud. 

\begin{figure}[t]
    \centering
    \includegraphics[width=\linewidth]{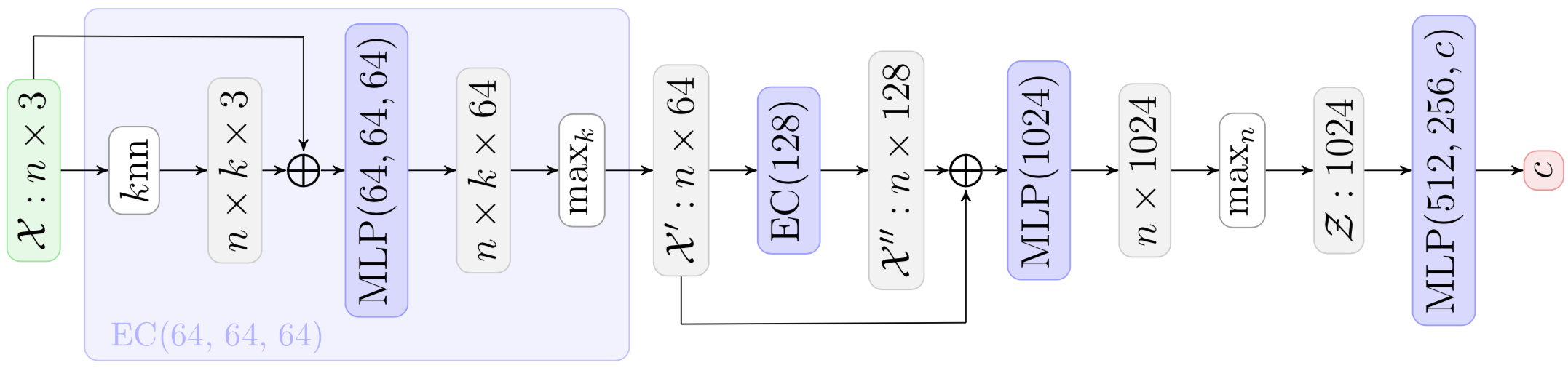}
    \caption{The DEC architecture adopted. Green, gray, and red blocks represent input, intermediate, and output tensors, respectively. Parametric layers are colored in blue, while fixed layers in white.}
    \label{fig:dec_arch}
\end{figure}

\begin{figure}[t]
    \centering
        \begin{subfigure}[t]{0.49\linewidth}
        \centering
        \includegraphics[width=0.53\linewidth]{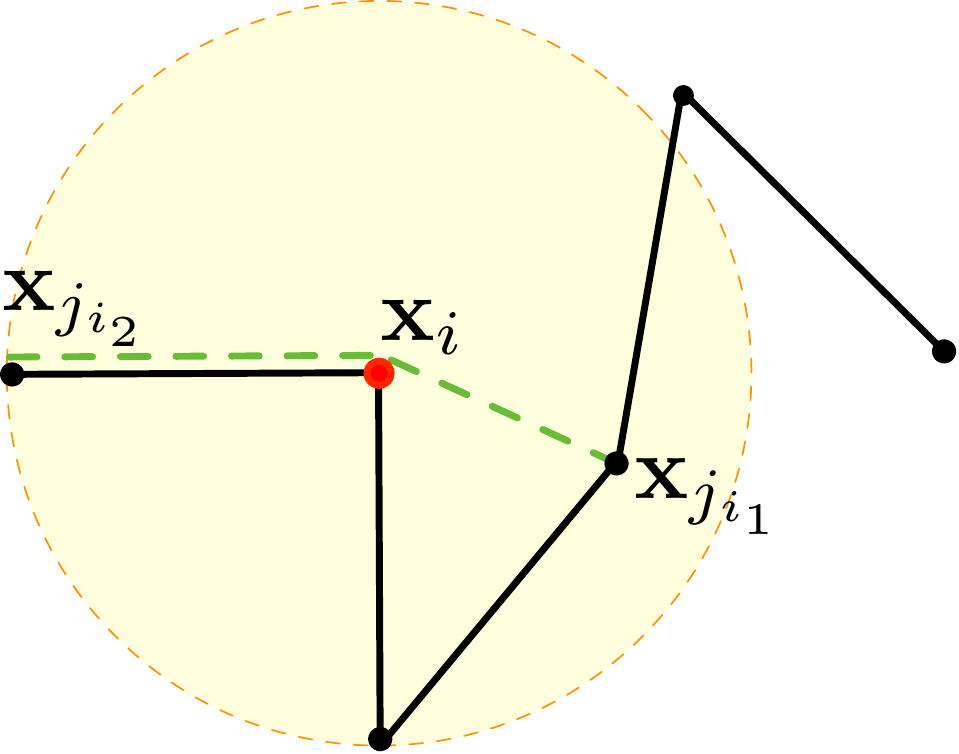}
        \caption{}
        \label{fig:knn_a}
    \end{subfigure}
    \begin{subfigure}[t]{0.49\linewidth}
        \centering
        \includegraphics[width=0.5\linewidth]{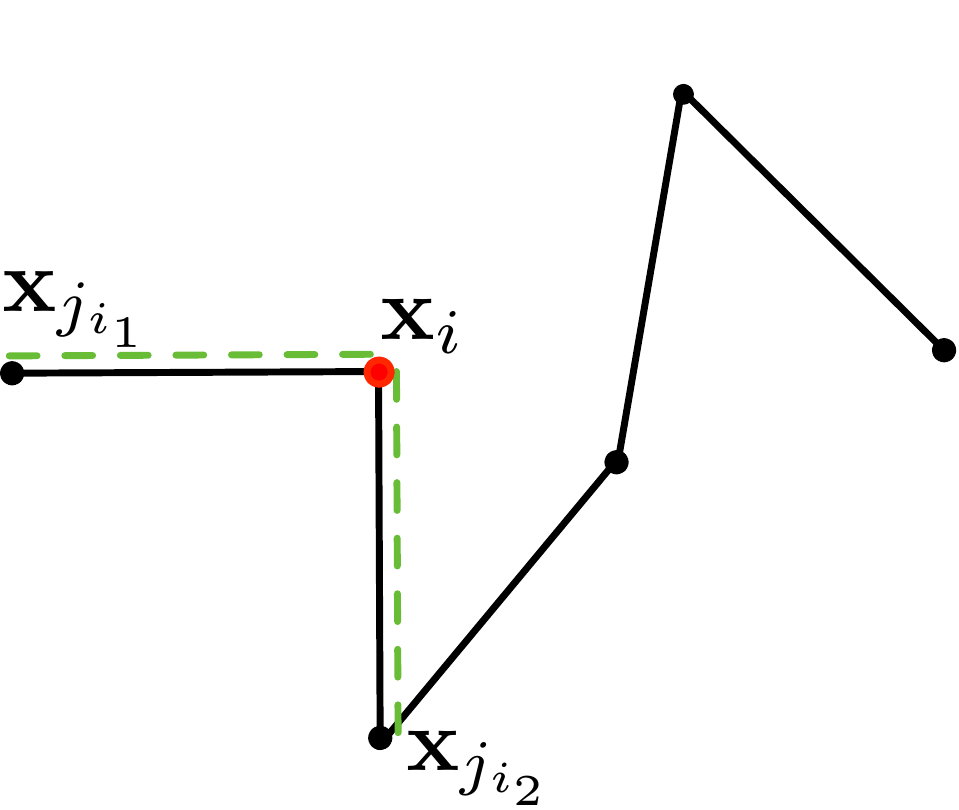}
        \caption{}
        \label{fig:knn_b}
    \end{subfigure}
    \caption{Comparison between Euclidean $k$-nn (a), and $k$-nn based on the sequence graph (b).}
\end{figure}

\paragraph{Dynamic Edge Convolution.} Considering a point cloud $\mathcal{X} = \{ \textbf{x}_1, \textbf{x}_2, \dots, \textbf{x}_n\}$, $\textbf{x}_i \in \mathbb{R}^3$, the DEC model first guesses an initial graph structure for the point cloud by retrieving for each point $\textbf{x}_i$ the set of $k$ nearest neighbors, $\mathrm{knn}(\textbf{x}_i) = \{\textbf{x}_{j_{i_1}}, \dots, \textbf{x}_{j_{i_k}}\}$, in terms of Euclidean distance (see Figure \ref{fig:knn_a}). Hence, DEC builds a $k$-nn graph, $\mathcal{G(V,E)}$, where $\mathcal{V}=\mathcal{X}$ is the set of nodes, and an edge $(i, j) \in \mathcal{E}$ exists iff $\textbf{x}_j \in \mathrm{knn}(\textbf{x}_i)$. Then, each point representation, $\textbf{x}_i$, is enriched with the representation of each of its neighbors $\textbf{x}_{j_i}$, creating edges features $\textbf{e}_{ij}$, which are learnt through a neural network $h_{\boldsymbol{\mathrm{\Theta}}}$, i.e. $\textbf{e}_{ij}=h_{\boldsymbol{\mathrm{\Theta}}}(\textbf{x}_i \oplus (\textbf{x}_j - \textbf{x}_i))$, where $\oplus$ denotes the concatenation operator. Finally, a new representation of a point, $\textbf{x}_i^\prime$, is obtained by aggregating all the learned edge features with a pooling operator, i.e. $\textbf{x}_i^{\prime}={\mathrm{pool}}(\textbf{e}_{ij})$, $j \colon (i,j) \in \mathcal{E}$, where $\mathrm{pool}$ is either $\mathrm{max}$ or $\mathrm{mean}$. The sequence of operations that from $\mathcal{X}$ produce $\mathcal{X}^{\prime}$ define an Edge Convolution (EC) layer, which in the DEC model is repeated multiple times. Referring to our architecture, see Figure \ref{fig:dec_arch}, two EC layers are stacked in depth to produce new representations $\mathcal{X}^{\prime}$ and $\mathcal{X}^{\prime\prime}$ with 64 and 128 features respectively. Each of the two EC layers computes its own $k$-nn graph in order to adjust the local neighborhood of points to its input representation, i.e, $\mathcal{X}$ and $\mathcal{X}^{\prime}$. The re-computation of knn is what defines the DEC model as \emph{dynamic}. Finally, the different learned representations are concatenated, encoded with a learning layer, $g_{\boldsymbol{\mathrm{\Phi}}}$, to 1024 features, and pooled to obtain a single descriptor of the point cloud, $\mathcal{Z}=\mathrm{pool}(g_{\boldsymbol{\mathrm{\Phi}}}(\mathcal{X}^{\prime} \oplus \mathcal{X}^{\prime\prime}))$, which is classified using a fully connected (FC) network.

\paragraph{Dynamic Edge Convolution of a Sequence.}
A remarkable property of the DEC model (also shared by PN) is the invariance to the permutation of the points in the input point cloud. Indeed, such models make use only of operators invariant to the order e.g., FC layers, max / mean pooling layers. Although this property is fundamental in the point cloud domain, it becomes undesired if the input is a sequence as in our case. To solve this issue, we propose a simple but well-motivated modification of the DEC model, in which we impose the input point cloud to have a graph structure, without needing a Euclidean $k$-nn to induce it. According to the streamline structure, we impose the input to be a bidirectional sequence graph where each non-terminal point, $\textbf{x}_{i \neq 0,n}$, has two neighbors: the previous and the next point in the sequence, while the terminal points, $\textbf{x}_0, \textbf{x}_n$, have just one neighbor (see Figure \ref{fig:knn_b}):
\[
\mathcal{G(V,E^\prime)},\, e^\prime_{ij} \in \mathcal{E^\prime} \colon \textbf{x}_i \to \textbf{x}_j,\, j=i+1 \vee j=i-1.
\]
It is important to remark that we impose this structure only in the first sDEC layer, which is enough to lose the invariance with respect to the input permutations while maintaining the invariance with respect to the flipping of the sequence, thanks to the bidirectionality of the sequence graph. Also, imposing this structure only in the first layer preserves the dynamicity of the model, which remains able to re-organize the structure of the point cloud in the latent space according to the task at hand. 


\section{Experiments and Results}%
\label{sec:experiments}

The main purpose of the experiments is to perform an empirical assessment of the behaviour of different models for the task of tractogram filtering. In our comparison, we consider a reference model, bLSTM, and three competing models based on geometric deep learning: PN, DEC, and sDEC.

Before setting the training of learning models, we carried out a few preprocessing steps. The first step was the pruning of fibers. An analysis of the distribution of fibers between P and nP according to the ExTractor’s rules, highlighted a massive percentage of very short streamlines, i.e. length below $20$ mm. All of them were labeled as nP. To reduce this potential bias for the learning models we removed such streamlines from the tractograms. After the pruning, the average distribution of P and nP classes was $68\%$ and $32\%$ respectively.
The second preprocessing step concerned the resampling of the points of streamlines. Traditional learning models, e.g. bLSTM, require input to be represented as fixed vectorial representations. Therefore, despite GDL models can deal with a varying number of points, we have to resample all streamlines to have the same number of points \cite{garyfallidis_quickbundles_2012,odonnell_automatic_2007,gupta_fibernet_2017}. We computed the resampling using a cubic B-spline interpolation and empirically compared different representations based on 12, 16 and 20 points. No meaningful performance differences were noticed. In Table~\ref{tab:result} we report the results for streamlines resampled to 16 points.

To perform a fair comparison of the four models, we have configured all of them to have a similar number of parameters, approximately around $900$K. While the architecture of DEC and sDEC models is reported in Figure~\ref{fig:dec_arch}, the configuration of the remaining models is: \\
\normalsize bLSTM:\quad \footnotesize \texttt{MLP(128) $\rightarrow$ LSTM(256) $\oplus$ LSTM$^{-1}$(256) $\rightarrow$ MLP(256, 128) $\rightarrow$ FC(2)};\\
\normalsize PN:\quad \footnotesize \texttt{MLP(64,64,64,128,1024) $\rightarrow$ MAX $\rightarrow$ MLP(512,256,40) $\rightarrow$ FC(2)},\\ \normalsize
where \texttt{\small FC} is a Fully Connected layer, and \texttt{\small MLP} contains sequences of (\texttt{\small FC, BatchNorm, ReLU}). For the implementation of the models, we adopted the PyTorch library \cite{paszke_pytorch_2019} with the PyTorch Geometric extension \cite{fey_fast_2019}. 

\newcolumntype{C}{>{\centering\arraybackslash}p{2cm}}
\begin{table}[t]
    \centering
    \begin{tabular}{ l | C | C | C | C }
        \hline
        Method & Accuracy & Precision & Recall & DSC \\
        \hline
        bLSTM   & 92.8 $(\pm0.3)$ & 93.7 $(\pm0.5)$  & 96.1 $(\pm0.2)$ & 94.9 $(\pm0.3)$ \\
        PN      & 94.5 $(\pm0.1)$ & 95.4 $(\pm0.2)$ & 96.8 $(\pm0.2)$ & 96.1 $(\pm0.2)$ \\
        DEC     & 94.3 $(\pm0.1)$ & 95.4 $(\pm0.3)$ & 96.5 $(\pm0.2)$ & 95.9 $(\pm0.2)$ \\
        sDEC    & \textbf{95.2} $(\pm0.1)$ & \textbf{96.2} $(\pm0.3)$ & \textbf{96.9} $(\pm0.1)$ & \textbf{96.6} $(\pm0.2)$ \\
        \hline
    \end{tabular}
    \vspace{0.2cm}
    \caption{Average scores for the 4 HCP subjects of test set. Standard deviation among the 4 subjects is reported between brackets.  
    }
    \label{tab:result}
\end{table}

For all the experiments we considered a dataset of 20 tractograms from 20 different subjects. We partitioned the data into three sets: 12 tractograms for training the models, 4 tractograms for hyperparameters tuning and model selection, and 4 tractograms for testing. 
The training was designed as follows: 1000 epochs with evaluation step every 20 epochs; cross-entropy loss; Adam optimizer with default alfa and beta momentum $(0.9, 0.99)$; initial learning rate of $10^{-3}$ multiplied by a factor of $0.7$ every $90$ epochs until a minimum value of $5\cdot 10^{-5}$ is reached. In the training procedure, we used mini-batches composed of $16$K streamlines randomly sampled from two subjects, $8$K from each of them. A subject is sampled only once for each epoch.
On the test set, we computed the following evaluation measures: accuracy, precision, recall, and DSC. In Table~\ref{tab:result} the values refer to the average for each subject. Notice that the classification task is single-streamline, therefore the size of our test set is composed of $\sim 3$ million streamlines.

\begin{figure}[t]
\includegraphics[width=1.0\textwidth]{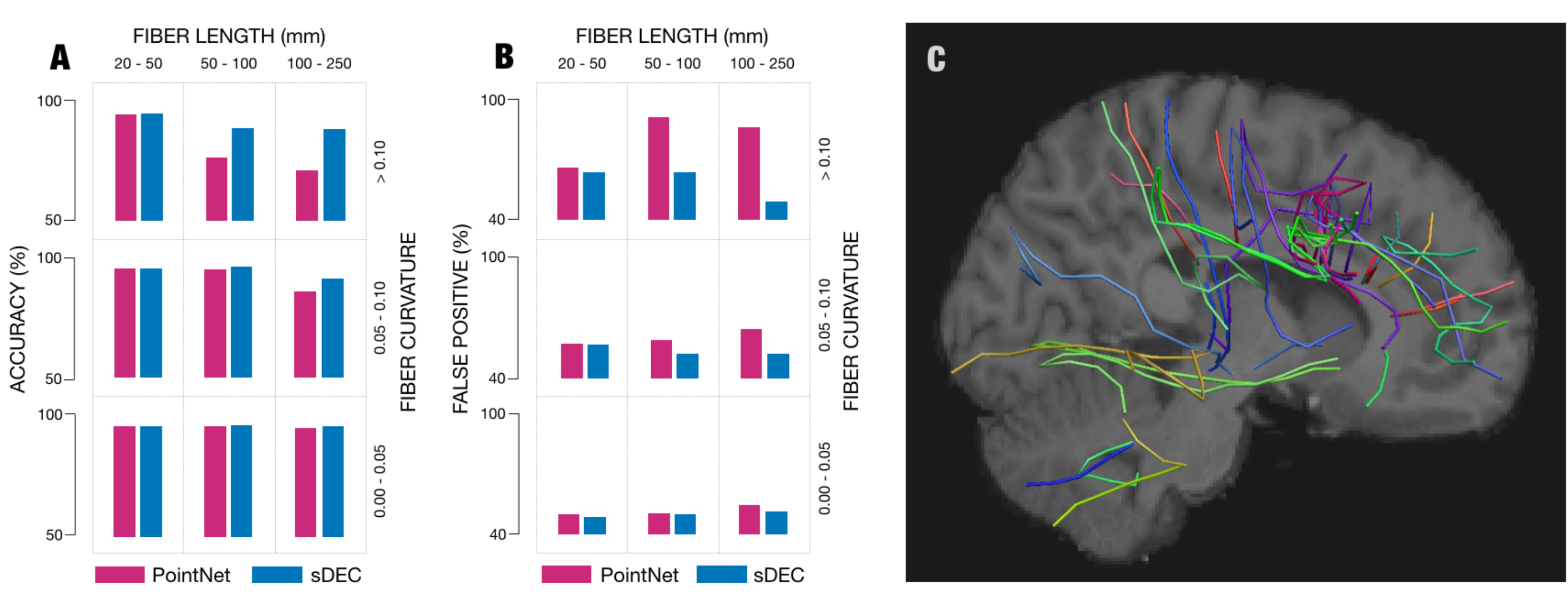} 
\caption{Analysis of the error distribution for PN and sDEC models with respect to streamlines length and curvature: (A) accuracy, (B) false positive rate. (C) Sample of false negative streamlines, mistakenly classified as anatomically non-plausible.}
\label{fig:error}
\end{figure}

An additional experiment was designed to test the invariance of DEC and sensitivity of sDEC with respect to the order of points in a streamline. To this end, we tested the two models on a version of the test set where the the sequence of points of streamlines were randomly permuted.  In agreement with the working hypothesis, the accuracy for DEC remained the same, $94.3\%$, while there was a drop for sDEC to $30.0\%$. 

A further analysis concerned the examination of the error distribution. We restricted the analysis to PN and sDEC only. Our interest was to characterize whether and how the error differs between the two models. We considered a couple of features in the analysis: length and curvature of streamlines. All streamlines were partitioned into three categories with respect to length (short $[0, 50]$ mm, medium $[50, 100]$ mm, long $[100, 300]$ mm) and curvature (straight $[0.0, 0.05]$, curved $[0.05, 0.10]$, very curved $[0.10, 0.20]$ ), where each of them contained at least a portion of $15\%$ of streamlines. Combining the partitions we obtained 9 categories. In Figure~\ref{fig:error}A we depict how the accuracy score varies across the categories. The task becomes more difficult when both length and curvature are greater. A similar analysis was carried out focusing on the distribution of false positive error, i.e. streamlines mistakenly classified as anatomically plausible. The results are reported in Figure~\ref{fig:error}B. In this case, we observe a major difference between the two models in the most critical categories, namely long and curved streamlines. In Figure~\ref{fig:error}C we illustrate a qualitative example of false negative errors, when streamlines should be recognized as anatomically plausible while they are classified as anatomically non-plausible. Visual inspection by an expert anatomist confirms that the classifier was indeed correct in those cases and that the labelling process was noisy.

\paragraph{Code and Data.} For the sake of reproducibility, we provide both the code and the dataset used in our experiments. We adopt the BrainLife\footnote{\url{https://brainlife.io/}} \cite{avesani_open_2019} platform to distribute them. The dataset containing the 20 tractograms with the respective labelling is available at \url{https://doi.org/10.25663/brainlife.pub.13}. For the code we distribute a pre-trained implementation of sDEC as a BrainLife app at \url{https://doi.org/10.25663/brainlife.app.390}, while the entire source code can be found at \url{https://github.com/FBK-NILab/tractogram_filtering/tree/miccai2020}.


\section{Discussion and Conclusions}%
\label{sec:discussion}

The results reported in Table~\ref{tab:result} confirm that we may successfully approach the task of tractogram filtering as a supervised learning problem. The best accuracy achieved by sDEC is beyond $95\%$ with low standard deviation across subjects. This performance is obtained considering only the representation of streamlines as a sequence of 3D points. The runtime application of sDEC model is very fast. We may filter a whole tractogram with $\sim 1M$ streamlines in less than one minute (46.2 sec using a gpu NVIDIA Titan Xp 12Gb).

We argue that the accuracy is underestimated and the true score may be even higher. If we look at the portion of false negative error, as reported in Figure~\ref{fig:error}C, we may agree that some of those streamline trajectories can be considered anatomically non-plausible, even though the true label states the contrary. Therefore, the computation of accuracy is biased by this apparent mismatch, which we can consider as part of the noise in the labelling process. Moreover, we may claim that the sDEC model has a good generalization capability and behaves accurately even in the presence of misleading labels.


The geometric deep learning models seem to provide only a small, even though meaningful (t-test with p-value $< 0.001$), improvement in terms of accuracy compared to bLSTM. Nevertheless PN and sDEC support higher flexibility in the streamlines representation, both as points clouds or graphs. The sDEC model seems the most appropriate for tractogram filtering because it is robust with respect to the order of points in the streamlines while preserving the highest accuracy in the discrimination of anatomical plausibility.

The analysis of error distribution allows a better understanding of the difference between PN and sDEC models. Around $40\%$ of the misclassification error concerns different streamlines, indicating that the two models behave differently: in Figure~\ref{fig:error}A the most critical streamlines are those longer and more curved. Nevertheless, sDEC is more robust and the drop in accuracy is lower than PN. We may explain this difference with the property of sDEC to capture the relations among the points because it is considering also the edges of streamlines. When streamlines are long and curved, edges become more informative and provide a competitive advantage. As illustrated in Figure~\ref{fig:error}B, PN has the bias to classify long and curved streamlines as anatomically plausible doing more false positive error.

We believe that a fast and accurate filtering of tractograms, like the one supported by the sDEC model, is the premise for further investigations on how tasks like bundle segmentation may take advantage of the removal of artifactual streamlines.

\end{document}